\documentclass[twocolumn,prl,aps,showpacs, superscriptaddress]{revtex4-1}

\usepackage{graphicx}
\usepackage{epstopdf}

\usepackage{pdfcomment}
\usepackage{color}

\begin{document}

\title{Hysteretic melting transition of a soliton lattice in a commensurate charge modulation}

\author{Pin-Jui Hsu} 
\affiliation{Physikalisches Institut, Universit{\"a}t W{\"u}rzburg, 97074 W{\"u}rzburg, Germany}
\author{Tobias Mauerer} 
\affiliation{Physikalisches Institut, Universit{\"a}t W{\"u}rzburg, 97074 W{\"u}rzburg, Germany}
\author{Matthias Vogt} 
\affiliation{Physikalisches Institut, Universit{\"a}t W{\"u}rzburg, 97074 W{\"u}rzburg, Germany}
\author{J.J. Yang}
\affiliation{Laboratory for Pohang Emergent Materials and Department of Physics, 
Pohang University of Science and Technology, Pohang 790-784, Republic of Korea}
\author{Yoon Seok Oh}
\affiliation{Rutgers Center for Emergent Materials and Department of Physics and Astronomy, 
Rutgers University, Piscataway, New Jersey 08854, USA}
\author{S-W. Cheong}
\affiliation{Laboratory for Pohang Emergent Materials and Department of Physics, 
Pohang University of Science and Technology, Pohang 790-784, Republic of Korea}
\affiliation{Rutgers Center for Emergent Materials and Department of Physics and Astronomy, 
Rutgers University, Piscataway, New Jersey 08854, USA}
\author{Matthias Bode}
\affiliation{Physikalisches Institut, Universit{\"a}t W{\"u}rzburg, 97074 W{\"u}rzburg, Germany}
\author{Weida Wu}
\email{wdwu@physics.rutgers.edu}
\affiliation{Rutgers Center for Emergent Materials and Department of Physics and Astronomy, 
Rutgers University, Piscataway, New Jersey 08854, USA}

\today

\begin{abstract}
We report on the observation of the hysteretic transition of a commensurate charge modulation in IrTe$_2$ from transport and scanning tunneling microscopy (STM) studies. Below the transition ($T_{\rm C} \approx 275$\,K on cooling) a $q = 1/5$ charge modulation was observed, which is consistent with previous studies. Additional modulations [$q_n = (3n+2)^{-1}$] appear below a second transition at  $T_{\rm S}\approx 180$\,K on cooling. The coexistence of various modulations persist up to $T_{\rm C}$ on warming.  The atomic structures of charge modulations and the temperature dependent STM studies suggest that 1/5 modulation is a periodic soliton lattice which partially melts below $T_{\rm S}$ on cooling. Our results provide compelling evidence that the ground state of IrTe$_2$ is a commensurate 1/6 charge modulation, which originates from periodic dimerization of Te atoms visualized by atomically resolved STM images. 
\end{abstract}
\maketitle
\newpage
A periodic electronic charge modulation, often called charge-density wave (CDW), is an ordering phenomenon which is accompanied by a distortion of the underlying lattice with the same periodicity\,\cite{gruner88}. 
The driving force of a CDW can be a Fermi surface nesting instability (also called Peierls transition; often considered  the canonical mechanism of charge or spin density waves in low dimensional electronic systems\,\cite{gruner88}), 
the indirect Jahn-Teller effect\,\cite{jahn37}, or the formation of local bound states\,\cite{jerome67, wezel10}. 
Often the CDW periodicity is incommensurate with respect to the underlying lattice. The competition between these two periodicities may lead to a transition from an incommensurate phase to a specific commensurate phase, i.e.\ a lock-in transition below the CDW phase transition\,\cite{bhatt75, mcmillan76, chen81, chen82}.  

An elegant picture of the lock-in transition is a melting transition of a soliton lattice with no soliton in the commensurate phase but a finite density of periodic solitons in the incommensurate phase\,\cite{bhatt75, mcmillan76}. 
While the commensurate phase is energetically favored at low temperature because of lower elastic energy cost and positive soliton energy, 
the incommensurate phase is favored by the driving mechanism (e.g.\ the nesting condition) and the entropy gain due to the formation of the soliton lattice\,\cite{mcmillan76,bak80}. 
The continuous reduction of the soliton density results in a continuous variation of the incommensurability of the charge modulation, 
which is closely related to the famous devil's staircase, where infinite commensurate phases 
with modulations at all possible fractional numbers emerge from competing microscopic interactions\,\cite{bak80, villain80, bak82, selke88}. 

Recently, an intriguing charge/orbital density wave was discovered in the 5d transition metal dichalcogenide IrTe$_2$ with large spin-orbital coupling\,\cite{yang12, pyon12}. 
Interestingly, superconductivity emerges with the suppression of the CDW phase\,\cite{yang12, pyon12, fang13}, indicating that the coexistence of CDW and superconductivity in quasi-2-dimensional (Q2D) transition metal dichalcogenides is a general phenomenon. 
Originally, Fermi surface nesting has been suggested as the mechanism of the charge modulation in IrTe$_2$\,\cite{yang12}. 
However, various experimental observations indicate that the local bonding states of Te orbitals and the mixed valence nature of Ir ions are more likely responsible for the driving force of the charge modulation\,\cite{fang13, oh13, ootsuki12}. 
Therefore, it is imperative to visualize the local electronic modulation with atomic resolution to reveal the fundamental mechanism of charge modulation in IrTe$_2$.  

Herein, we report on the observation of an intriguing hysteretic transition ($T_{\rm S}\approx 180$\,K) of the commensurate charge modulation phase in IrTe$_2$ below the previously reported transition ($T_{\rm C}\approx 275$\,K) using transport and STM measurements. 
The atomically resolved modulations suggest that the previously observed 1/5 modulation is a periodic lattice of solitonlike phase slips, similar to the unidirectional CDW due to periodic phase slips observed in strained NbSe$_2$\,\cite{soumy13}.  
In this picture, the hysteretic transition may be described as an incomplete melting of a soliton lattice with a sudden decrease of soliton density, resulting in short-range charge modulations described by $q_n = (3n+2)^{-1}$. 
On warming the partially melted soliton lattice persists all the way to $T_{\rm C}$. 
The partial melting transition and the coexistence of multiple fractional modulations are similar to the characteristics of the devil's staircase phenomena\,\cite{bak80, villain80}.
Our high-resolution STM data suggest that the intrinsic ground state of IrTe$_2$ is likely an 1/6 modulation with periodic Te dimer stripes.   

\begin{figure}[t]
\includegraphics[width=\columnwidth]{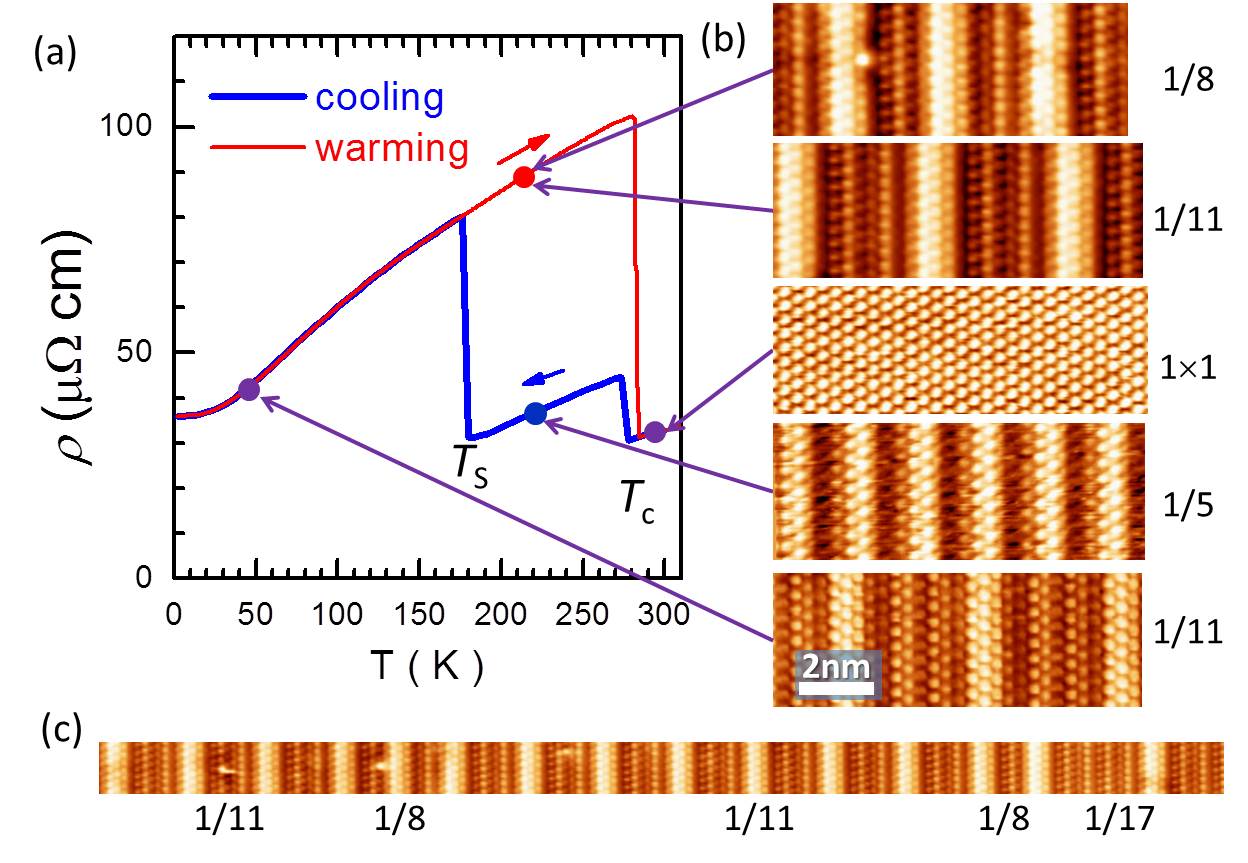}
\caption{(Color online) (a) $\rho(T)$ data of a IrTe$_2$ single crystal. 
The CDW transition ($T_{\rm C} \approx 275$\,K) is indicated by the first resistivity increase. 
The second resistivity increase at $\sim$ 180\,K indicates the partial melting transition ($T_{\rm S}$) on cooling.  
(b) STM topographic images with atomic resolution taken at 295, 220, 50 (cooling) and 215 K (warming).  
The various new modulations (1/8, 1/11,  etc.) are present below $T_{\rm S}$. 
(c) STM image (65\,nm$\times$\,3\,nm) at 215\,K (warming) showing an example of coexistence of several modulations: 1/8, 1/11 and 1/17. 
\label{fig1} }
\end{figure}

Single crystalline IrTe$_{2}$ specimens were grown from the Te flux. 
STM experiments were performed in an ultra-high vacuum (UHV) multifunctional chamber with a base pressure $p \le 1\times10^{-10}$\,mbar. 
The single crystals were cleaved in UHV at room temperature (RT) to expose the pristine (001) surface, presumably terminated by the Te layer. 
STM images were acquired in constant-current mode with typical STM parameters of $|U| = 0.01\ ...\ 1$\,V and $I = 0.1\ ...\ 2$\,nA.  
Magnetization and electrical resistivity were measured up to 400\,K using the Quantum Design MPMS-XL7 and PPMS-9. STM images were analyzed with WSXM \cite{wsxm}. 

Figure \ref{fig1}(a) shows the typical $\rho(T)$ data measured on a high quality single crystalline IrTe$_2$ sample. 
In addition to the CDW transition ($T_{\rm C}$) reported in earlier works
\,\cite{yang12, pyon12, fang13, oh13}, another transition can be recognized on cooling the sample at $T_{\rm S}\approx 180$\,K which is also characterized by a resistivity increase. 
On warming, the high-resistivity state persists to $T_{\rm C}$ without returning to the intermediate state. This unusually large hysteresis was reproduced in multiple single crystals, indicative of the significant metastability of the low temperature state. 
The absence of the second transition ($T_{\rm S}$) in earlier works suggests that it is very sensitive to the sample quality. In fact, tiny inhomogeneous strain is sufficient to smear out the sharp $\rho(T)$ anomalies\,\footnote{See supplemental material.}.

Previous scattering studies suggest the CDW transition ($T_{\rm C}$) is characterized by formation of a $\vec{q}=1/5(1,0,\bar{1})$ superstructure\,\cite{yang12, oh13}. 
This is directly confirmed by our STM observation of 1/5(1,0) on the (001) surface at intermediate temperature (220\,K) on cooling, while the RT-STM image in Fig.\,\ref{fig1}(b) shows the expected 1$\times$1 atomic structure. Therefore, our STM observation of 1/5 modulation unlikely originates from surface reconstruction.\,\cite{Note1}

\begin{figure}[t]
\includegraphics[width=\columnwidth]{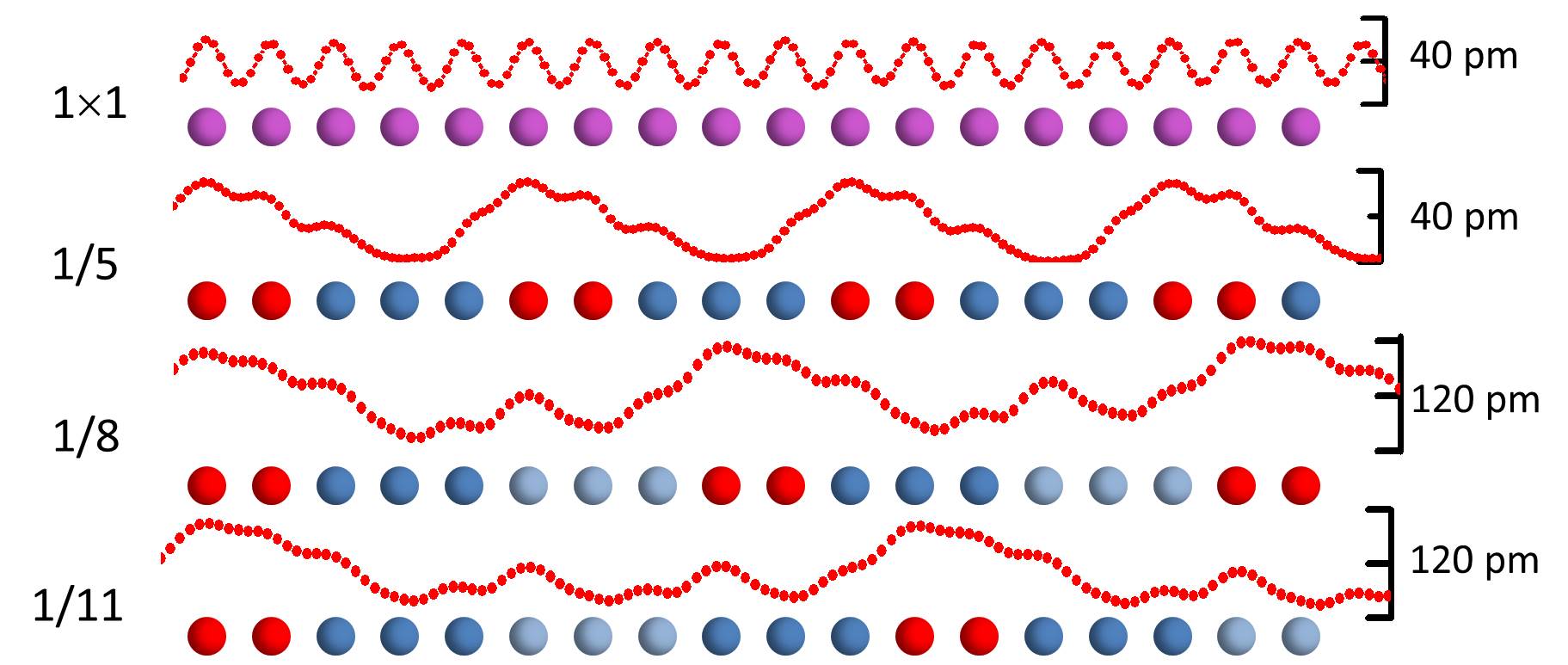}
\caption{(Color online) Column-averaged line profiles of STM images in Fig.\,\ref{fig1} and  the cartoon models of multiple modulations $q_n$=1/5, 1/8 and 1/11 and RT undistorted state.
Here the double-columns (red spheres) are the soliton-like (anti-phase) boundaries of fundamental modulation of three-rows (blue spheres). 
\label{fig2} }
\end{figure}

At $T < T_{\rm S}$ new periodicities (e.g. 1/8, 1/11, etc.) with domain sizes of a few nanometers appear. They consist of double-columns (appearing bright in color-coded STM images) which are separated by an integer number of three-row blocks, as shown in Fig.\,\ref{fig1}.  
The two types of fundamental units, 2$a^*$ and 3$a^*$ atomic columns, are clearly visible in column-averaged line profiles 
 which have been extracted from the atomically resolved STM images in Fig.\ref{fig1} and are plotted as red traces in Fig.\,\ref{fig2}. Here $a^*=a\sin60^{\circ}\approx3.4$\,\AA\ is the inter-column spacing, and $a$ is lattice constant. 
The composition for 2$a^*$ and 3$a^*$ units is highlighted by red and blue spheres, respectively. 
Thereby, our STM results suggest that the new modulations can be unified in a simple empirical relationship, namely, $\lambda_n=(3n+2)a^*$ with $n\ge1$. This picture suggests that the ground state of IrTe$_2$ is probably a periodic structure of 3$a^*$ columns. In this sense, a 2$a^*$ column corresponds to a fractional phase slip, i.e.\ a soliton-like (anti-phase) boundary of the ground state\,\cite{zhang11, soumy13}.
Within this picture the 1/5 modulation is a soliton lattice with maximal density, and the hysteretic transition $T_{\rm S}$ is a partial ``melting'' of the soliton lattice. 
This scenario is very similar to the soliton lattice melting picture of the lock-in transition in Q2D CDW systems\,\cite{mcmillan76, chen81, chen82}. 

\begin{figure*}[t]
\includegraphics[width=.95\textwidth]{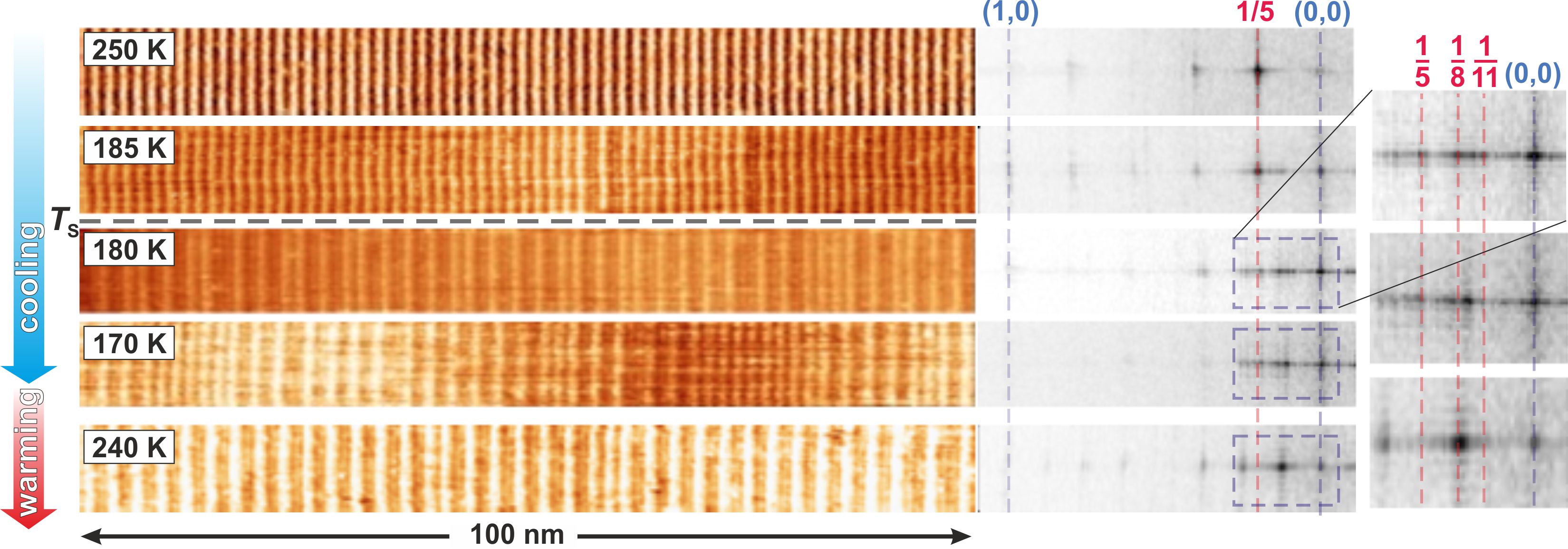}
\caption{(Color online) STM topographic images (left) and their FFT maps (right) 
across the $T_{\rm S}$ on cooling (250, 185, 180 and 170\,K) and on warming (240\,K). 
The density of solitions (bright stripes) decreases across $T_{\rm S}$
with appearance of short-range modulations of $q_n=(3n+2)^{-1}$.  
The sharp 1/5 superlattice spots turn into diffusive streaks because of the random packing of these short-range modulations.\,\cite{Note1} 
On warming (240\,K) the 1/8 superlattice spot becomes sharper.
\label{fig3}}
\end{figure*}

To reveal the mechanism of the second transition, we performed variable temperature STM studies of IrTe$_2$ single crystals between 250\,K and 140\,K. 
Because of the lack of unique topographic features\,\cite{Note1} it is difficult to keep track of the same surface area by compensating the thermal drift after each temperature change (unlike in previous studies\,\cite{hsu13}). 
To circumvent this difficulty, large-scale STM images with atomic resolution are needed for statistical analysis of the evolution of modulations. 
For this purpose, multiple ($>5$) STM images with a width of 200\,nm along the modulation direction were taken at various temperatures. 
Representative sections of STM images at 250, 185, 180, 170 (cooling) and 240\,K (warming) and their corresponding fast Fourier transformed (FFT) images are shown in Fig.\,\ref{fig3}. 
Clearly a uniform 1/5 modulation is observed in the intermediate temperature range  $T_{\rm S}<T<T_{\rm C}$ with correlation length over many microns, resulting in sharp 1/5 spots in FFT maps.
Similar to recent scattering studies, the second harmonic peak (2/5) is quite strong, suggesting a non-sinusoidal modulation\,\cite{kir13, cao13}. 
Interestingly, new modulations of longer wavelengths but very short correlation lengths appear below $T_{\rm S}$ as shown in STM images, which is consistent with the diffusive superlattice streaks\,\cite{Note1}. 
The short correlation lengths suggest that the new modulations probably nucleate and get trapped inside the original 1/5 domains. 
Consistently, negligible change of soliton density across $T_{\rm S}$ was observed near a twin domain boundary with two different $\vec{q}$ directions (not shown here), excluding the twin boundaries as possible nucleation centers of new modulations. 
On warming the correlation length of 1/8 increases substantially resulting in a sharp spot in FFT map (at 240\,K), indicative of  thermal annealing of the soliton lattice.

\begin{figure}[t]
\includegraphics[width=0.9\columnwidth]{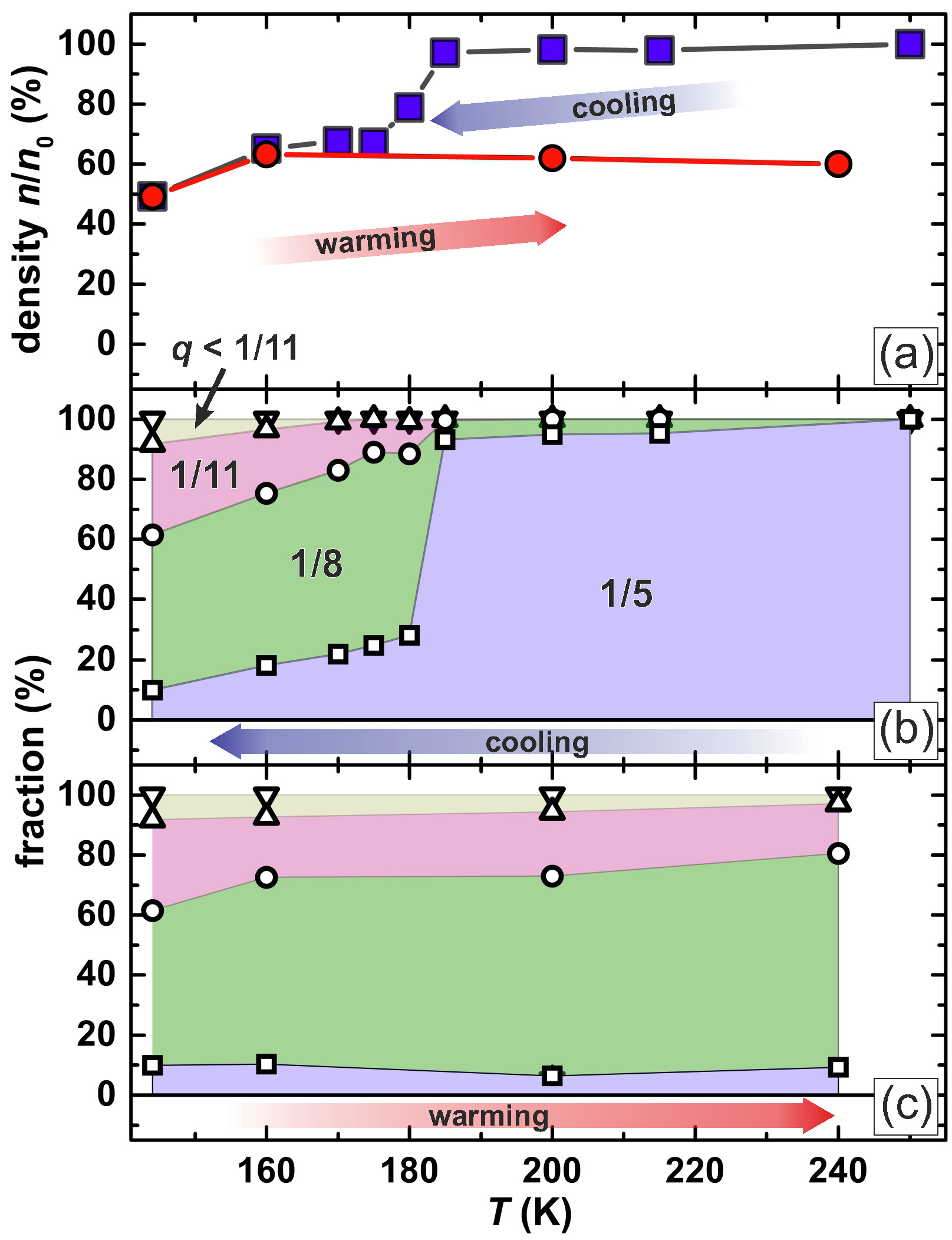}
\caption{$T$-dependence of the normalized soliton density $n/n_0$ across $T_{\rm S}$ (a) and areal fractions of various $q_n$'s on cooling (b) and warming (c).  
The statistical data are obtained from counting the bright 2$a^*$ columns in STM images with dimension $>200$ nm along the $\vec{q}_n$ direction.
\label{fig4} }
\end{figure}

Figure\,\ref{fig4} summarizes our counting statistics of STM observation\,\cite{Note1}. 
Fig.\,\ref{fig4}(a) shows the $T$-dependence of normalized soliton density $n/n_0$ (bright 2$a^*$ columns  in STM topographic images, such as in Fig.\,\ref{fig3})  where $n$ is the soliton density from data and $n_0$ is that of pure 1/5 modulation.  
Above $T_{\rm S}$, the soliton density is almost 100\%, while it drops significantly to $\sim 70\%$ across $T_{\rm S}$, which is in good agreement with transport data\,\cite{Note1}.
As $T$ decreases further, $n/n_0$ continues to decrease, indicating that solitons are energetically unfavorable in the low temperature phase.  
On warming $n/n_0$ stays at $\sim60\%$ up to 240\,K, in good agreement with the transport data in Fig.\,\ref{fig1}(a).
The anti-correlation between $n(T)$ and $\rho(T)$ suggests that the solitons (2$a^*$) has a lower resistivity than the ground state (3$a^*$). 
 In contrast to the significant increase of soliton lattice correlation length (of 1/8) on warming, the $\rho(T)$ has a similar slope as that of the intermediate phase (1/5), indicating that the defect scattering of conduction electrons is not sensitive to the disorders of the soliton lattice. 
Therefore, the sharp increase of resistivity at $T_{\rm S}$ is mainly caused by a reduction of density of states at Fermi level ($N_F$), indicating that solitons have a higher $N_F$ than the ground state. 
Consistently, the solitons appear higher in constant-current STM topography at low bias. 

More detailed information can be obtained from the areal fractions of individual modulations (1/5, 1/8, 1/11, etc.) shown in Fig.\,\ref{fig4}(b) and (c).  
The areal fractions of various modulations show similar hysteretic behaviors,  i.e., significant changes across $T_{\rm S}$ on cooling but minor changes on warming. 
Specifically, the fraction of 1/5 reduces drastically from over 90\% to $\sim 30\%$ across $T_{\rm S}$, while the fractions of other modulations (especially 1/8) increase substantially.

All of our STM data are consistent with the soliton lattice picture of commensurate charge modulations discussed in above text, which provides a simple unified description [$q_n=(3n+2)^{-1}$] of all the observed modulations in IrTe$_2$. 
The soliton lattice melting transition at $T_{\rm S}$ indicates that the ground state of IrTe$_2$ consists of a periodic structure of $3a^*$ stripes. 
Indeed, a abrupt change of modulation from 1/5 to 1/6 was observed in Se doped IrTe$_2$ where the transition temperature $T_{\rm C}$ systematically increases with increasing Se concentration\,\cite{oh13}. 
Because Se is more electronegative than Te, this observation provides compelling evidence that the CDW transition is likely caused by breaking Te-Te bonds\,\cite{oh13}. 
However, the origin of the 1/6 modulation is unclear in previous studies\,\cite{oh13}. 
Our STM results provide a natural explanation of the 1/5 to 1/6 transition in Se doped IrTe$_2$,  namely, that the weakening of Te-Te bonds by Se doping energetically favors the intrinsic ground state of 1/6 super-structure (i.e.\ a doubling of  $3a^*$ stripes). This scenario is supported by the observation of a sudden increase of the thermal hysteresis ($\Delta T$) of $T_{\rm C}$ at the boundary between 1/5 and 1/6\,\cite{oh13}. 

\begin{figure}[t]
\includegraphics[width=.9\columnwidth]{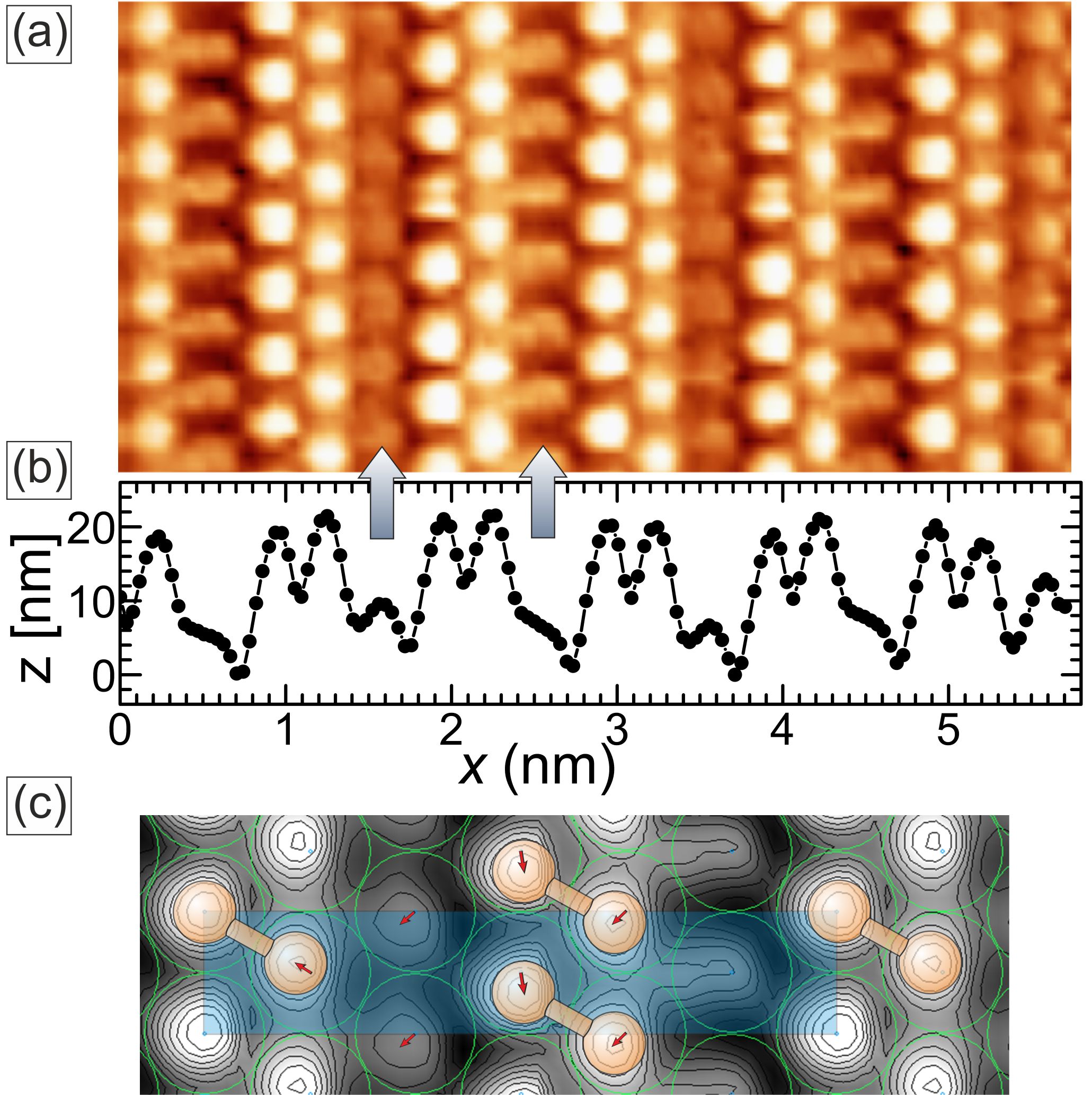}
\caption{(Color online) (a) An atomically-resolved STM image measured at 50\,K of a region without any soliton, and (b) its column-averaged line profile. The two bright atomic rows are $\sim0.4$\,\AA\ closer to each other than the undistorted inter-column spacing (3.4\,\AA), suggesting a formation of Te double-row stripes separated by unpaired Te atomic rows. There is a subtle difference between alternating unpaired Te atomic rows as noted by two arrows in (b), resulting in a $6a^*$ periodicity. (c) Unit cell average of the STM image in (a). The center box highlights the $1\times6$ supercell. Green circles label the undistorted hexagonal lattice. The atomic displacement of individual atoms is labeled by red arrows. See main text for numerical values. Yellow dumbbells highlight the Te dimers. 
\label{fig5} }
\end{figure}

Our picture is further corroborated by the occasional observation of large patches (dimension over 100\,nm) of purely 1/6 modulation (i.e.\ no soliton) below $T_{\rm S}$ in IrTe$_2$.  
A zoom-in STM image with atomic resolution of the 1/6 modulation and its corresponding line profile are shown in Fig.\,\ref{fig5}(a) and (b). 
There is a subtle difference between the neighboring $3a^*$ stripes, as indicated by the two arrows in Fig.\,\ref{fig5}. 
The microscopic origin of the two kind of $3a^*$ stripes is unclear at this moment.
The spacing between two bright atomic rows ($\sim$3.0\,\AA) is significantly smaller than the mean inter-column spacing (3.4\,\AA), indicating a  formation of Te double-rows separated by unpaired Te rows. 
Detailed analysis of atomic displacements of Te atoms within the $1\times6$ supercell [shown in Fig.\,\ref{fig5}(c)] indicates the formation of Te dimers within each Te double-row. 
The atomic displacements with respect to the undistorted $1\times1$ lattice are approximately (unit:\,\AA): $(0,0), (-0.4, 0.4), (-0.4, -0.5), (0, -0.6), (-0.4, -0.5)$, and $(0, 0)$, which are derived from column averaged profiles ($x$ displacements) and inter-column cross-correlations ($y$ displacements). 
The Te dimer stripes might be induced by the formation of Ir$^{4+}$ dimers proposed by high resolution x-ray scattering analysis\,\cite{kir13}. 

Previous studies suggested that the charge modulation transition originates from breaking of the Te-Te bonds because of mixed valence states of Ir (3+/4+)\,\cite{jobic92, ootsuki12, fang13, oh13}. Since there exists one hole in the $t_{2g}$ orbital of an Ir$^{4+}$ ion\,\cite{radaelli02}, it has been speculated that the charge modulation is associated with orbital rearrangement\,\cite{pyon12}, which is supported by XPS measurements and scattering refinements\,\cite{ootsuki12, cao13}.  Single crystal x-ray scattering refinement analyses suggest that both Ir and Te form periodic dimer stripes below $T_{\rm C}$\,\cite{kir13}, which is consistent with our STM observation.

In conclusion, we have carried our a systematic high resolution STM study of the charge modulated phases in high quality single crystals of IrTe$_2$ and discovered a hysteretic melting transition of soliton lattice that occurs only on cooling. 
Our results suggest that the previously reported 1/5 modulation is a periodic soliton lattice that partially melts at $T_{\rm S}\approx 180$\,K. 
Our STM observation provides compelling evidence that the ground state of IrTe$_2$ is an $1/6$ super-structure with periodic Te dimer stripes.  
The melting transition is interrupted probably by kinetic energy barriers due to the formation of a dilute soliton lattice with short-range modulations described by $q_n = (3n+2)^{-1}$.  

We thank G.L. Pascut and V. Kiryukhin for sharing unpublished x-ray results and the idea of dimerization. The work at Rutgers was supported by the NSF-DMREF-1233349 and DMR-0844807. 
The work at Postech was supported by the Max Planck POSTECH/KOREA Research Initiative Program [No. 2011-0031558] through the NRF of Korea funded by the Ministry of Education, Science and Technology. The work at W\"{u}rzburg was supported by Deutsche Forschungsgemeinschaft within FOR 1162 (grant BO1468/20-1) 
and by the Alexander von Humboldt Foundation (WW).

\bibliography{IrTe2}

\begin{thebibliography}{26}%
\makeatletter
\providecommand \@ifxundefined [1]{%
 \@ifx{#1\undefined}
}%
\providecommand \@ifnum [1]{%
 \ifnum #1\expandafter \@firstoftwo
 \else \expandafter \@secondoftwo
 \fi
}%
\providecommand \@ifx [1]{%
 \ifx #1\expandafter \@firstoftwo
 \else \expandafter \@secondoftwo
 \fi
}%
\providecommand \natexlab [1]{#1}%
\providecommand \enquote  [1]{``#1''}%
\providecommand \bibnamefont  [1]{#1}%
\providecommand \bibfnamefont [1]{#1}%
\providecommand \citenamefont [1]{#1}%
\providecommand \href@noop [0]{\@secondoftwo}%
\providecommand \href [0]{\begingroup \@sanitize@url \@href}%
\providecommand \@href[1]{\@@startlink{#1}\@@href}%
\providecommand \@@href[1]{\endgroup#1\@@endlink}%
\providecommand \@sanitize@url [0]{\catcode `\\12\catcode `\$12\catcode
  `\&12\catcode `\#12\catcode `\^12\catcode `\_12\catcode `\%12\relax}%
\providecommand \@@startlink[1]{}%
\providecommand \@@endlink[0]{}%
\providecommand \url  [0]{\begingroup\@sanitize@url \@url }%
\providecommand \@url [1]{\endgroup\@href {#1}{\urlprefix }}%
\providecommand \urlprefix  [0]{URL }%
\providecommand \Eprint [0]{\href }%
\providecommand \doibase [0]{http://dx.doi.org/}%
\providecommand \selectlanguage [0]{\@gobble}%
\providecommand \bibinfo  [0]{\@secondoftwo}%
\providecommand \bibfield  [0]{\@secondoftwo}%
\providecommand \translation [1]{[#1]}%
\providecommand \BibitemOpen [0]{}%
\providecommand \bibitemStop [0]{}%
\providecommand \bibitemNoStop [0]{.\EOS\space}%
\providecommand \EOS [0]{\spacefactor3000\relax}%
\providecommand \BibitemShut  [1]{\csname bibitem#1\endcsname}%
\let\auto@bib@innerbib\@empty
\bibitem [{\citenamefont {Gr\"uner}(1988)}]{gruner88}%
  \BibitemOpen
  \bibfield  {author} {\bibinfo {author} {\bibfnamefont {G.}~\bibnamefont
  {Gr\"uner}},\ }\href {\doibase 10.1103/RevModPhys.60.1129} {\bibfield
  {journal} {\bibinfo  {journal} {Rev. Mod. Phys.}\ }\textbf {\bibinfo {volume}
  {60}},\ \bibinfo {pages} {1129} (\bibinfo {year} {1988})}\BibitemShut
  {NoStop}%
\bibitem [{\citenamefont {Jahn}\ and\ \citenamefont {Teller}(1937)}]{jahn37}%
  \BibitemOpen
  \bibfield  {author} {\bibinfo {author} {\bibfnamefont {H.~A.}\ \bibnamefont
  {Jahn}}\ and\ \bibinfo {author} {\bibfnamefont {E.}~\bibnamefont {Teller}},\
  }\href@noop {} {\bibfield  {journal} {\bibinfo  {journal} {Proc. R. Soc.
  London A}\ }\textbf {\bibinfo {volume} {161}},\ \bibinfo {pages} {220}
  (\bibinfo {year} {1937})}\BibitemShut {NoStop}%
\bibitem [{\citenamefont {J\'erome}\ \emph {et~al.}(1967)\citenamefont
  {J\'erome}, \citenamefont {Rice},\ and\ \citenamefont {Kohn}}]{jerome67}%
  \BibitemOpen
  \bibfield  {author} {\bibinfo {author} {\bibfnamefont {D.}~\bibnamefont
  {J\'erome}}, \bibinfo {author} {\bibfnamefont {T.~M.}\ \bibnamefont {Rice}},
  \ and\ \bibinfo {author} {\bibfnamefont {W.}~\bibnamefont {Kohn}},\ }\href
  {\doibase 10.1103/PhysRev.158.462} {\bibfield  {journal} {\bibinfo  {journal}
  {Phys. Rev.}\ }\textbf {\bibinfo {volume} {158}},\ \bibinfo {pages} {462}
  (\bibinfo {year} {1967})}\BibitemShut {NoStop}%
\bibitem [{\citenamefont {van Wezel}\ and\ \citenamefont
  {Littlewood}(2010)}]{wezel10}%
  \BibitemOpen
  \bibfield  {author} {\bibinfo {author} {\bibfnamefont {J.}~\bibnamefont {van
  Wezel}}\ and\ \bibinfo {author} {\bibfnamefont {P.~B.}\ \bibnamefont
  {Littlewood}},\ }\href {\doibase 10.1103/Physics.3.87} {\bibfield  {journal}
  {\bibinfo  {journal} {Physics}\ }\textbf {\bibinfo {volume} {3}},\ \bibinfo
  {pages} {87} (\bibinfo {year} {2010})}\BibitemShut {NoStop}%
\bibitem [{\citenamefont {Bhatt}\ and\ \citenamefont
  {McMillan}(1975)}]{bhatt75}%
  \BibitemOpen
  \bibfield  {author} {\bibinfo {author} {\bibfnamefont {R.~N.}\ \bibnamefont
  {Bhatt}}\ and\ \bibinfo {author} {\bibfnamefont {W.~L.}\ \bibnamefont
  {McMillan}},\ }\href {\doibase 10.1103/PhysRevB.12.2042} {\bibfield
  {journal} {\bibinfo  {journal} {Phys. Rev. B}\ }\textbf {\bibinfo {volume}
  {12}},\ \bibinfo {pages} {2042} (\bibinfo {year} {1975})}\BibitemShut
  {NoStop}%
\bibitem [{\citenamefont {McMillan}(1976)}]{mcmillan76}%
  \BibitemOpen
  \bibfield  {author} {\bibinfo {author} {\bibfnamefont {W.~L.}\ \bibnamefont
  {McMillan}},\ }\href {\doibase 10.1103/PhysRevB.14.1496} {\bibfield
  {journal} {\bibinfo  {journal} {Phys. Rev. B}\ }\textbf {\bibinfo {volume}
  {14}},\ \bibinfo {pages} {1496} (\bibinfo {year} {1976})}\BibitemShut
  {NoStop}%
\bibitem [{\citenamefont {Chen}\ \emph {et~al.}(1981)\citenamefont {Chen},
  \citenamefont {Gibson},\ and\ \citenamefont {Fleming}}]{chen81}%
  \BibitemOpen
  \bibfield  {author} {\bibinfo {author} {\bibfnamefont {C.~H.}\ \bibnamefont
  {Chen}}, \bibinfo {author} {\bibfnamefont {J.~M.}\ \bibnamefont {Gibson}}, \
  and\ \bibinfo {author} {\bibfnamefont {R.~M.}\ \bibnamefont {Fleming}},\
  }\href {\doibase 10.1103/PhysRevLett.47.723} {\bibfield  {journal} {\bibinfo
  {journal} {Phys. Rev. Lett.}\ }\textbf {\bibinfo {volume} {47}},\ \bibinfo
  {pages} {723} (\bibinfo {year} {1981})}\BibitemShut {NoStop}%
\bibitem [{\citenamefont {Chen}\ \emph {et~al.}(1982)\citenamefont {Chen},
  \citenamefont {Gibson},\ and\ \citenamefont {Fleming}}]{chen82}%
  \BibitemOpen
  \bibfield  {author} {\bibinfo {author} {\bibfnamefont {C.~H.}\ \bibnamefont
  {Chen}}, \bibinfo {author} {\bibfnamefont {J.~M.}\ \bibnamefont {Gibson}}, \
  and\ \bibinfo {author} {\bibfnamefont {R.~M.}\ \bibnamefont {Fleming}},\
  }\href {\doibase 10.1103/PhysRevB.26.184} {\bibfield  {journal} {\bibinfo
  {journal} {Phys. Rev. B}\ }\textbf {\bibinfo {volume} {26}},\ \bibinfo
  {pages} {184} (\bibinfo {year} {1982})}\BibitemShut {NoStop}%
\bibitem [{\citenamefont {Bak}\ and\ \citenamefont {von Boehm}(1980)}]{bak80}%
  \BibitemOpen
  \bibfield  {author} {\bibinfo {author} {\bibfnamefont {P.}~\bibnamefont
  {Bak}}\ and\ \bibinfo {author} {\bibfnamefont {J.}~\bibnamefont {von
  Boehm}},\ }\href {\doibase 10.1103/PhysRevB.21.5297} {\bibfield  {journal}
  {\bibinfo  {journal} {Phys. Rev. B}\ }\textbf {\bibinfo {volume} {21}},\
  \bibinfo {pages} {5297} (\bibinfo {year} {1980})}\BibitemShut {NoStop}%
\bibitem [{\citenamefont {Villain}\ and\ \citenamefont
  {Gordon}(1980)}]{villain80}%
  \BibitemOpen
  \bibfield  {author} {\bibinfo {author} {\bibfnamefont {J.}~\bibnamefont
  {Villain}}\ and\ \bibinfo {author} {\bibfnamefont {M.~B.}\ \bibnamefont
  {Gordon}},\ }\href {\doibase 10.1088/0022-3719/13/17/005} {\bibfield
  {journal} {\bibinfo  {journal} {J. Phys. C: Solid State Phys.}\ }\textbf
  {\bibinfo {volume} {13}},\ \bibinfo {pages} {3117} (\bibinfo {year}
  {1980})}\BibitemShut {NoStop}%
\bibitem [{\citenamefont {Bak}\ and\ \citenamefont {Bruinsma}(1982)}]{bak82}%
  \BibitemOpen
  \bibfield  {author} {\bibinfo {author} {\bibfnamefont {P.}~\bibnamefont
  {Bak}}\ and\ \bibinfo {author} {\bibfnamefont {R.}~\bibnamefont {Bruinsma}},\
  }\href {\doibase 10.1103/PhysRevLett.49.249} {\bibfield  {journal} {\bibinfo
  {journal} {Phys. Rev. Lett.}\ }\textbf {\bibinfo {volume} {49}},\ \bibinfo
  {pages} {249} (\bibinfo {year} {1982})}\BibitemShut {NoStop}%
\bibitem [{\citenamefont {Selke}(1988)}]{selke88}%
  \BibitemOpen
  \bibfield  {author} {\bibinfo {author} {\bibfnamefont {W.}~\bibnamefont
  {Selke}},\ }\href {\doibase http://dx.doi.org/10.1016/0370-1573(88)90140-8}
  {\bibfield  {journal} {\bibinfo  {journal} {Physics Reports}\ }\textbf
  {\bibinfo {volume} {170}},\ \bibinfo {pages} {213 } (\bibinfo {year}
  {1988})}\BibitemShut {NoStop}%
\bibitem [{\citenamefont {Yang}\ \emph {et~al.}(2012)\citenamefont {Yang},
  \citenamefont {Choi}, \citenamefont {Oh}, \citenamefont {Hogan},
  \citenamefont {Horibe}, \citenamefont {Kim}, \citenamefont {Min},\ and\
  \citenamefont {Cheong}}]{yang12}%
  \BibitemOpen
  \bibfield  {author} {\bibinfo {author} {\bibfnamefont {J.~J.}\ \bibnamefont
  {Yang}}, \bibinfo {author} {\bibfnamefont {Y.~J.}\ \bibnamefont {Choi}},
  \bibinfo {author} {\bibfnamefont {Y.~S.}\ \bibnamefont {Oh}}, \bibinfo
  {author} {\bibfnamefont {A.}~\bibnamefont {Hogan}}, \bibinfo {author}
  {\bibfnamefont {Y.}~\bibnamefont {Horibe}}, \bibinfo {author} {\bibfnamefont
  {K.}~\bibnamefont {Kim}}, \bibinfo {author} {\bibfnamefont {B.~I.}\
  \bibnamefont {Min}}, \ and\ \bibinfo {author} {\bibfnamefont {S.-W.}\
  \bibnamefont {Cheong}},\ }\href {\doibase 10.1103/PhysRevLett.108.116402}
  {\bibfield  {journal} {\bibinfo  {journal} {Phys. Rev. Lett.}\ }\textbf
  {\bibinfo {volume} {108}},\ \bibinfo {pages} {116402} (\bibinfo {year}
  {2012})}\BibitemShut {NoStop}%
\bibitem [{\citenamefont {Pyon}\ \emph {et~al.}(2012)\citenamefont {Pyon},
  \citenamefont {Kudo},\ and\ \citenamefont {Nohara}}]{pyon12}%
  \BibitemOpen
  \bibfield  {author} {\bibinfo {author} {\bibfnamefont {S.}~\bibnamefont
  {Pyon}}, \bibinfo {author} {\bibfnamefont {K.}~\bibnamefont {Kudo}}, \ and\
  \bibinfo {author} {\bibfnamefont {M.}~\bibnamefont {Nohara}},\ }\href
  {\doibase 10.1143/JPSJ.81.053701} {\bibfield  {journal} {\bibinfo  {journal}
  {J. Phys. Soc. Jap.}\ }\textbf {\bibinfo {volume} {81}},\ \bibinfo {pages}
  {053701} (\bibinfo {year} {2012})}\BibitemShut {NoStop}%
\bibitem [{\citenamefont {Fang}\ \emph {et~al.}(2013)\citenamefont {Fang},
  \citenamefont {Xu}, \citenamefont {Dong}, \citenamefont {Zheng},\ and\
  \citenamefont {Wang}}]{fang13}%
  \BibitemOpen
  \bibfield  {author} {\bibinfo {author} {\bibfnamefont {A.~F.}\ \bibnamefont
  {Fang}}, \bibinfo {author} {\bibfnamefont {G.}~\bibnamefont {Xu}}, \bibinfo
  {author} {\bibfnamefont {T.}~\bibnamefont {Dong}}, \bibinfo {author}
  {\bibfnamefont {P.}~\bibnamefont {Zheng}}, \ and\ \bibinfo {author}
  {\bibfnamefont {N.~L.}\ \bibnamefont {Wang}},\ }\href {\doibase
  10.1038/srep01153} {\bibfield  {journal} {\bibinfo  {journal} {Sci. Rep.}\
  }\textbf {\bibinfo {volume} {3}},\ \bibinfo {pages} {1153} (\bibinfo {year}
  {2013})}\BibitemShut {NoStop}%
\bibitem [{\citenamefont {Oh}\ \emph {et~al.}(2013)\citenamefont {Oh},
  \citenamefont {Yang}, \citenamefont {Horibe},\ and\ \citenamefont
  {Cheong}}]{oh13}%
  \BibitemOpen
  \bibfield  {author} {\bibinfo {author} {\bibfnamefont {Y.~S.}\ \bibnamefont
  {Oh}}, \bibinfo {author} {\bibfnamefont {J.~J.}\ \bibnamefont {Yang}},
  \bibinfo {author} {\bibfnamefont {Y.}~\bibnamefont {Horibe}}, \ and\ \bibinfo
  {author} {\bibfnamefont {S.-W.}\ \bibnamefont {Cheong}},\ }\href {\doibase
  10.1103/PhysRevLett.110.127209} {\bibfield  {journal} {\bibinfo  {journal}
  {Phys. Rev. Lett.}\ }\textbf {\bibinfo {volume} {110}},\ \bibinfo {pages}
  {127209} (\bibinfo {year} {2013})}\BibitemShut {NoStop}%
\bibitem [{\citenamefont {Ootsuki}\ \emph {et~al.}(2012)\citenamefont
  {Ootsuki}, \citenamefont {Wakisaka}, \citenamefont {Pyon}, \citenamefont
  {Kudo}, \citenamefont {Nohara}, \citenamefont {Arita}, \citenamefont {Anzai},
  \citenamefont {Namatame}, \citenamefont {Taniguchi}, \citenamefont {Saini},\
  and\ \citenamefont {Mizokawa}}]{ootsuki12}%
  \BibitemOpen
  \bibfield  {author} {\bibinfo {author} {\bibfnamefont {D.}~\bibnamefont
  {Ootsuki}}, \bibinfo {author} {\bibfnamefont {Y.}~\bibnamefont {Wakisaka}},
  \bibinfo {author} {\bibfnamefont {S.}~\bibnamefont {Pyon}}, \bibinfo {author}
  {\bibfnamefont {K.}~\bibnamefont {Kudo}}, \bibinfo {author} {\bibfnamefont
  {M.}~\bibnamefont {Nohara}}, \bibinfo {author} {\bibfnamefont
  {M.}~\bibnamefont {Arita}}, \bibinfo {author} {\bibfnamefont
  {H.}~\bibnamefont {Anzai}}, \bibinfo {author} {\bibfnamefont
  {H.}~\bibnamefont {Namatame}}, \bibinfo {author} {\bibfnamefont
  {M.}~\bibnamefont {Taniguchi}}, \bibinfo {author} {\bibfnamefont {N.~L.}\
  \bibnamefont {Saini}}, \ and\ \bibinfo {author} {\bibfnamefont
  {T.}~\bibnamefont {Mizokawa}},\ }\href {\doibase 10.1103/PhysRevB.86.014519}
  {\bibfield  {journal} {\bibinfo  {journal} {Phys. Rev. B}\ }\textbf {\bibinfo
  {volume} {86}},\ \bibinfo {pages} {014519} (\bibinfo {year}
  {2012})}\BibitemShut {NoStop}%
\bibitem [{\citenamefont {Soumyanarayanan}\ \emph {et~al.}(2013)\citenamefont
  {Soumyanarayanan}, \citenamefont {Yee}, \citenamefont {He}, \citenamefont
  {van Wezel}, \citenamefont {Rahn}, \citenamefont {Rossnagel}, \citenamefont
  {Hudson}, \citenamefont {Norman},\ and\ \citenamefont {Hoffman}}]{soumy13}%
  \BibitemOpen
  \bibfield  {author} {\bibinfo {author} {\bibfnamefont {A.}~\bibnamefont
  {Soumyanarayanan}}, \bibinfo {author} {\bibfnamefont {M.~M.}\ \bibnamefont
  {Yee}}, \bibinfo {author} {\bibfnamefont {Y.}~\bibnamefont {He}}, \bibinfo
  {author} {\bibfnamefont {J.}~\bibnamefont {van Wezel}}, \bibinfo {author}
  {\bibfnamefont {D.~J.}\ \bibnamefont {Rahn}}, \bibinfo {author}
  {\bibfnamefont {K.}~\bibnamefont {Rossnagel}}, \bibinfo {author}
  {\bibfnamefont {E.~W.}\ \bibnamefont {Hudson}}, \bibinfo {author}
  {\bibfnamefont {M.~R.}\ \bibnamefont {Norman}}, \ and\ \bibinfo {author}
  {\bibfnamefont {J.~E.}\ \bibnamefont {Hoffman}},\ }\href {\doibase
  10.1073/pnas.1211387110} {\bibfield  {journal} {\bibinfo  {journal} {PNAS}\
  }\textbf {\bibinfo {volume} {110}},\ \bibinfo {pages} {1623} (\bibinfo {year}
  {2013})}\BibitemShut {NoStop}%
\bibitem [{\citenamefont {Horcas}\ \emph {et~al.}(2007)\citenamefont {Horcas},
  \citenamefont {Fernandez}, \citenamefont {Gomez-Rodriguez}, \citenamefont
  {Colchero}, \citenamefont {Gomez-Herrero},\ and\ \citenamefont
  {Baro}}]{wsxm}%
  \BibitemOpen
  \bibfield  {author} {\bibinfo {author} {\bibfnamefont {I.}~\bibnamefont
  {Horcas}}, \bibinfo {author} {\bibfnamefont {R.}~\bibnamefont {Fernandez}},
  \bibinfo {author} {\bibfnamefont {J.~M.}\ \bibnamefont {Gomez-Rodriguez}},
  \bibinfo {author} {\bibfnamefont {J.}~\bibnamefont {Colchero}}, \bibinfo
  {author} {\bibfnamefont {J.}~\bibnamefont {Gomez-Herrero}}, \ and\ \bibinfo
  {author} {\bibfnamefont {A.~M.}\ \bibnamefont {Baro}},\ }\href {\doibase
  10.1063/1.2432410} {\bibfield  {journal} {\bibinfo  {journal} {Rev. Sci.
  Instrum.}\ }\textbf {\bibinfo {volume} {78}},\ \bibinfo {eid} {013705}
  (\bibinfo {year} {2007})}\BibitemShut {NoStop}%
\bibitem [{Note1()}]{Note1}%
  \BibitemOpen
  \bibinfo {note} {See supplemental material.}\BibitemShut {Stop}%
\bibitem [{\citenamefont {Zhang}\ \emph {et~al.}(2011)\citenamefont {Zhang},
  \citenamefont {Choi}, \citenamefont {Xu}, \citenamefont {Wang}, \citenamefont
  {Zhai}, \citenamefont {Wang}, \citenamefont {Zeng}, \citenamefont {Cho},
  \citenamefont {Zhang},\ and\ \citenamefont {Hou}}]{zhang11}%
  \BibitemOpen
  \bibfield  {author} {\bibinfo {author} {\bibfnamefont {H.}~\bibnamefont
  {Zhang}}, \bibinfo {author} {\bibfnamefont {J.-H.}\ \bibnamefont {Choi}},
  \bibinfo {author} {\bibfnamefont {Y.}~\bibnamefont {Xu}}, \bibinfo {author}
  {\bibfnamefont {X.}~\bibnamefont {Wang}}, \bibinfo {author} {\bibfnamefont
  {X.}~\bibnamefont {Zhai}}, \bibinfo {author} {\bibfnamefont {B.}~\bibnamefont
  {Wang}}, \bibinfo {author} {\bibfnamefont {C.}~\bibnamefont {Zeng}}, \bibinfo
  {author} {\bibfnamefont {J.-H.}\ \bibnamefont {Cho}}, \bibinfo {author}
  {\bibfnamefont {Z.}~\bibnamefont {Zhang}}, \ and\ \bibinfo {author}
  {\bibfnamefont {J.~G.}\ \bibnamefont {Hou}},\ }\href {\doibase
  10.1103/PhysRevLett.106.026801} {\bibfield  {journal} {\bibinfo  {journal}
  {Phys. Rev. Lett.}\ }\textbf {\bibinfo {volume} {106}},\ \bibinfo {pages}
  {026801} (\bibinfo {year} {2011})}\BibitemShut {NoStop}%
\bibitem [{\citenamefont {Hsu}\ \emph {et~al.}(2013)\citenamefont {Hsu},
  \citenamefont {Mauerer}, \citenamefont {Wu},\ and\ \citenamefont
  {Bode}}]{hsu13}%
  \BibitemOpen
  \bibfield  {author} {\bibinfo {author} {\bibfnamefont {P.-J.}\ \bibnamefont
  {Hsu}}, \bibinfo {author} {\bibfnamefont {T.}~\bibnamefont {Mauerer}},
  \bibinfo {author} {\bibfnamefont {W.}~\bibnamefont {Wu}}, \ and\ \bibinfo
  {author} {\bibfnamefont {M.}~\bibnamefont {Bode}},\ }\href {\doibase
  10.1103/PhysRevB.87.115437} {\bibfield  {journal} {\bibinfo  {journal} {Phys.
  Rev. B}\ }\textbf {\bibinfo {volume} {87}},\ \bibinfo {pages} {115437}
  (\bibinfo {year} {2013})}\BibitemShut {NoStop}%
\bibitem [{\citenamefont {Pascut}\ and\ \citenamefont
  {Kiryukhin}(2013)}]{kir13}%
  \BibitemOpen
  \bibfield  {author} {\bibinfo {author} {\bibfnamefont {G.~L.}\ \bibnamefont
  {Pascut}}\ and\ \bibinfo {author} {\bibfnamefont {V.}~\bibnamefont
  {Kiryukhin}},\ }\href@noop {} {}\bibinfo {howpublished} {private
  communication} (\bibinfo {year} {2013})\BibitemShut {NoStop}%
\bibitem [{\citenamefont {Cao}\ \emph {et~al.}(2013)\citenamefont {Cao},
  \citenamefont {Chakoumakos}, \citenamefont {Chen}, \citenamefont {Yan},
  \citenamefont {McGuire}, \citenamefont {Yang}, \citenamefont {Custelcean},
  \citenamefont {Zhou}, \citenamefont {Singh},\ and\ \citenamefont
  {Mandrus}}]{cao13}%
  \BibitemOpen
  \bibfield  {author} {\bibinfo {author} {\bibfnamefont {H.}~\bibnamefont
  {Cao}}, \bibinfo {author} {\bibfnamefont {B.~C.}\ \bibnamefont
  {Chakoumakos}}, \bibinfo {author} {\bibfnamefont {X.}~\bibnamefont {Chen}},
  \bibinfo {author} {\bibfnamefont {J.}~\bibnamefont {Yan}}, \bibinfo {author}
  {\bibfnamefont {M.~A.}\ \bibnamefont {McGuire}}, \bibinfo {author}
  {\bibfnamefont {H.}~\bibnamefont {Yang}}, \bibinfo {author} {\bibfnamefont
  {R.}~\bibnamefont {Custelcean}}, \bibinfo {author} {\bibfnamefont
  {H.}~\bibnamefont {Zhou}}, \bibinfo {author} {\bibfnamefont {D.~J.}\
  \bibnamefont {Singh}}, \ and\ \bibinfo {author} {\bibfnamefont
  {D.}~\bibnamefont {Mandrus}},\ }\href@noop {} {\  (\bibinfo {year} {2013})},\
  \Eprint {http://arxiv.org/abs/1302.5369} {arXiv:1302.5369} \BibitemShut
  {NoStop}%
\bibitem [{\citenamefont {Jobic}\ \emph {et~al.}(1992)\citenamefont {Jobic},
  \citenamefont {Brec},\ and\ \citenamefont {Rouxel}}]{jobic92}%
  \BibitemOpen
  \bibfield  {author} {\bibinfo {author} {\bibfnamefont {S.}~\bibnamefont
  {Jobic}}, \bibinfo {author} {\bibfnamefont {R.}~\bibnamefont {Brec}}, \ and\
  \bibinfo {author} {\bibfnamefont {J.}~\bibnamefont {Rouxel}},\ }\href
  {\doibase http://dx.doi.org/10.1016/S0022-4596(05)80309-3} {\bibfield
  {journal} {\bibinfo  {journal} {J. Solid State Chem.}\ }\textbf {\bibinfo
  {volume} {96}},\ \bibinfo {pages} {169 } (\bibinfo {year}
  {1992})}\BibitemShut {NoStop}%
\bibitem [{\citenamefont {Radaelli}\ \emph {et~al.}(2002)\citenamefont
  {Radaelli}, \citenamefont {Horibe}, \citenamefont {Gutmann}, \citenamefont
  {Ishibashi}, \citenamefont {Chen}, \citenamefont {Ibberson}, \citenamefont
  {Koyama}, \citenamefont {Hor}, \citenamefont {Kiryukhin},\ and\ \citenamefont
  {Cheong}}]{radaelli02}%
  \BibitemOpen
  \bibfield  {author} {\bibinfo {author} {\bibfnamefont {P.~G.}\ \bibnamefont
  {Radaelli}}, \bibinfo {author} {\bibfnamefont {Y.}~\bibnamefont {Horibe}},
  \bibinfo {author} {\bibfnamefont {M.~J.}\ \bibnamefont {Gutmann}}, \bibinfo
  {author} {\bibfnamefont {H.}~\bibnamefont {Ishibashi}}, \bibinfo {author}
  {\bibfnamefont {C.~H.}\ \bibnamefont {Chen}}, \bibinfo {author}
  {\bibfnamefont {R.~M.}\ \bibnamefont {Ibberson}}, \bibinfo {author}
  {\bibfnamefont {Y.}~\bibnamefont {Koyama}}, \bibinfo {author} {\bibfnamefont
  {Y.-S.}\ \bibnamefont {Hor}}, \bibinfo {author} {\bibfnamefont
  {V.}~\bibnamefont {Kiryukhin}}, \ and\ \bibinfo {author} {\bibfnamefont
  {S.-W.}\ \bibnamefont {Cheong}},\ }\href {\doibase 10.1038/416155a}
  {\bibfield  {journal} {\bibinfo  {journal} {Nature}\ }\textbf {\bibinfo
  {volume} {416}},\ \bibinfo {pages} {155} (\bibinfo {year}
  {2002})}\BibitemShut {NoStop}%
\end{thebibliography}%

\end{document}